\documentclass[prl,aps,twocolumn,superscriptaddress,showpacs,10pt,floatfix]{revtex4}
\usepackage{amsmath}
\usepackage{amsfonts}
\usepackage{amssymb}
\usepackage{bm}
\usepackage{graphicx}
\usepackage{datetime}
\usepackage{color}

\begin{document}
\newcommand{\figwidth}{0.95\columnwidth}
\newcommand{\ffigwidth}{0.4\columnwidth}
\newcommand{\warwick}{Department of Physics and Centre for Scientific Computing, University of Warwick, Coventry, CV4 7AL, United Kingdom}
\newcommand{\mainz}{Institut f\"{u}r Physik, Universit\"{a}t Mainz, D-55099 Mainz, Germany and Physik-Department E13, Technische Universit\"{a}t M\"{u}nchen,
D-85747 Garching, Germany}
\title{Anderson universality in a model of disordered phonons}
\author{Sebastian D. Pinski}\affiliation{\warwick}
\email[Corresponding author: ]{s.d.pinski@warwick.ac.uk}
\author{Walter Schirmacher}\affiliation{\mainz}
\author{Rudolf A. R\"omer}\affiliation{\warwick} 
\date{$Revision: 1.42 $, compiled \today, \currenttime}
%
\begin{abstract}
We consider the localization properties of a lattice of coupled masses and springs with random mass and spring constant values. We establish the full phase diagrams of the system for pure mass and pure spring disorder. The phase diagrams exhibit regions of stable as well as unstable  wave modes. The latter are of interest for the instantaneous-normal-mode spectra of liquids and the nascent field of acoustic metamaterials. We show the existence of delocalization-localization transitions throughout the phase diagram and establish, by high-precision numerical studies, that the universality of these transitions is of the Anderson type. 
\end{abstract}
\maketitle


Coherent wave phenomena in disordered systems are a recurring theme of modern physics. For condensed matter at the quantum scale, Anderson localization \cite{And58,EveM08} has recently been re-modernised by a series of beautiful experiments for Bose-Einstein condensates and the exponential decay of the localized waves has been directly measured \cite{BilJZB08}.
Results agree very well with recent spatially-resolved studies in semiconductor systems \cite{MorKMG02}. 
Similarly, light localization continues to be at the forefront of many research activities \cite{WieBLR97}. 
For classical waves, localization phenomena have an equally impressive history \cite{PagHST07} 
and recent ultra-sound propagation experiments \cite{HuSPS08} 
can now probe the spatial and multifractal structure of states close to the Anderson transition.
Phonon localization, i.e. absence of diffusion of acoustic or vibrational degrees of freedom has been addressed first in the seminal paper by John et al. \cite{JSS83} and thereafter in connection with the low-temperature thermal properties of glasses and the related enhancement of the vibrational density of states, the so-called ``boson peak'' \cite{AkkMay85,SchiWag93,SchDG98,KBS98,LudSTE01}. Also the localization properties of the Hessian Matrix of the potential-energy landscape of liquids and glasses --- ``instantaneous normal modes''  --- have been investigated and related to the liquid-glass transition \cite{Keyes94}.


A very interesting new avenue of research has opened up recently due to the realisation that the seminal work on electromagnetic metamaterials \cite{Ves68} has a companion in acoustic systems as well \cite{LiuZMZ00}. Hence hitherto unexplored and deemed unphysical regions of the phase diagram for disordered vibrations --- those with apparently \emph{negative} masses and spring constants --- are now recognised to be of considerable interest for metamaterial applications and offer an entirely novel perspective of Anderson localization. 

In this paper, we address the localization properties of a simple cubic lattice of particles with varying mass $m_i$ and varying nearest-neighbour harmonic force constants $k_{ij}$ . We present for the first time the complete phonon localization diagram for such a system, including the unstable regime. We achieve this by using high-precision transfer-matrix methods (TMM).

For simplicity we deal with ``scalar displacements'' $u_i(t)=u_i(\omega)e^{i\omega t}$ \cite{SchDG98} obeying equations of motions in the frequency domain
\begin{equation}\label{eq-dynmat}
-\omega^2m_iu_i=\sum_jk_{ij}(u_j-u_i)\, ,
\end{equation}
where the sum runs over the 6 nearest neighbours of a lattice site $i$.

In our calculations we use two types of disorder: In the case of {\it mass disorder} we allow the masses $m_i$ to be uniformly distributed in the interval $[ \overline{m}-\Delta m/2, \overline{m}+\Delta m/2]$ with $\overline{m}=1$ and keep $k_{ij}=\overline{k}=1$ constant. In the case of {\it spring disorder} we keep $m_i=\overline{m}=1$ constant and distribute the $k_{ij}$ uniformly in the interval $[1-\Delta k/2, 1+\Delta k/2]$. 
If the width of the distributions $\Delta m$ and $\Delta k$ exceeds $2$, negative masses/spring constants appear and mimic the unstable part of the Hessian of a liquid or the harmonic properties of an acoustic meta-material.

The present model can be related to Anderson's \cite{And58} tight-binding model for electron localization with Hamiltonian ${\cal H}=\sum_i|i\rangle\epsilon_i\langle i|+\sum_{ij}|i\rangle t_{ij}\langle j|$. 
For {\it mass disorder} we set $t_{ij} \equiv k_{ij}=const.=1$ and obtain the relations ($E$ is the quantum energy)
\begin{equation}\label{eq-puremass}
 E \equiv -\omega^2 + 6  \qquad 
 \epsilon_i \equiv \omega^2 \left( m_i-1 \right), 
\end{equation}
while for \emph{spring disorder} ($m_i=\overline{m}=1$), we have
\begin{equation}
\label{eq-purespring} 
 E \equiv -\omega^2  + 6  \qquad t_{ij}\,\equiv\,k_{ij}\qquad
 \epsilon_i \equiv -\sum_jk_{ij}.
\end{equation}
\begin{figure*}[tb]
(a)\includegraphics[width=\figwidth]{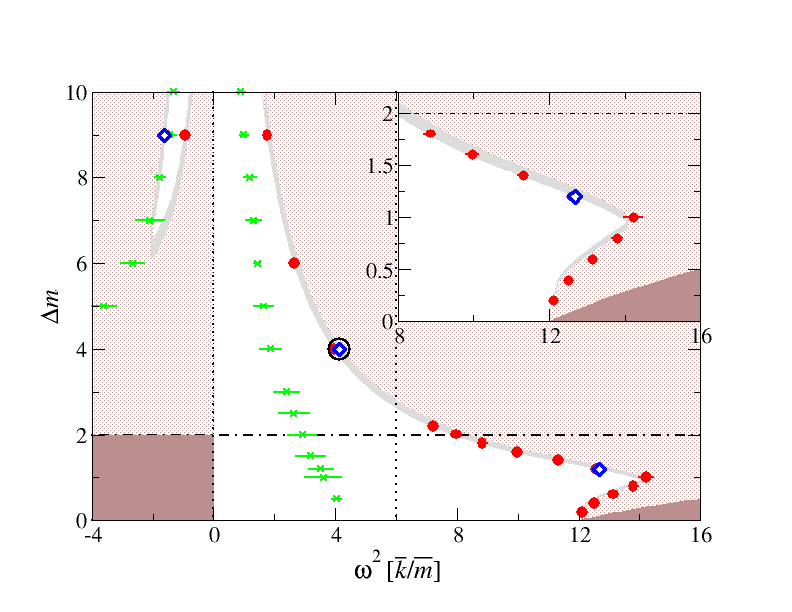}
(b)\includegraphics[width=\figwidth]{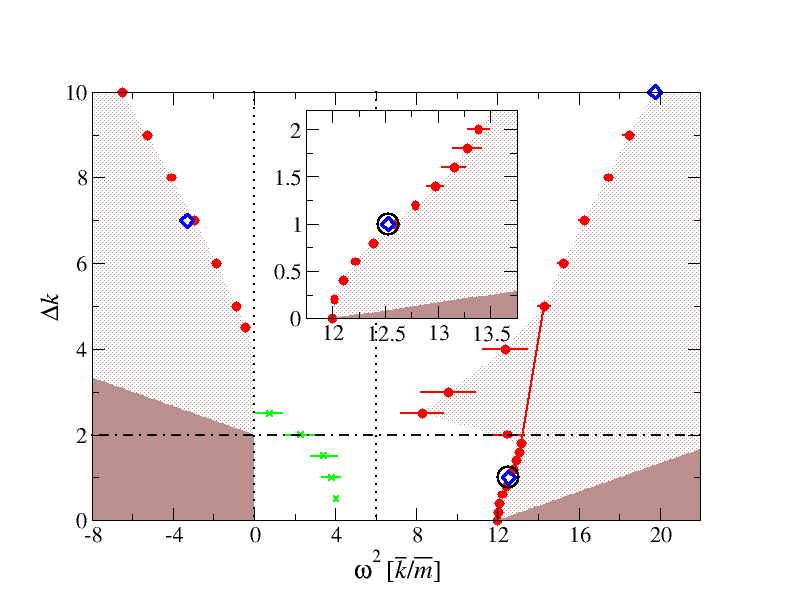}
\caption{(Color online) Phase diagrams of (a) mass disorder with disorder parameter $\Delta m$ and (b) spring disorder with disorder parameter $\Delta k$ versus squared frequency $\omega^2$. Shaded lines in (a) denote the critical region obtained from the transformation of the electronic phase diagram of \cite{BulSK87} via Eqs.\ (\ref{eq-puremass}). Light shadings denote localized regions, dark shadings regions beyond the band edges. Open blue diamonds denote transition point determined by finite-size scaling. Solid red circles denote estimated transition points (see text) and green crosses denote maxima in the DOS, divided by $\omega^2$ (``boson peaks''). The large open circles denote the locations of the critical eigenstates in Fig.\ \ref{fig-states}. The horizontal dash-dotted lines indicates the border between stable and unstable regions, the dotted lines denote $\omega^2=0$ and $6$. The red line for (b) marks a region of low numerical accuracy where the phase boundary is less well determined. Error bars for TMM data as described in the text. For the boson peak, errors denote width of the peak at $95\%$ of its height.
Insets: blow-ups of the phase boundaries in the stable regions.}
\label{phase}
\end{figure*}
With the relations \eqref{eq-puremass} and \eqref{eq-purespring} we can reuse many of the results for the Anderson model \cite{BulSK87,GruS95}. In particular, the analogy establishes the existence of localization-delocalization transitions for the vibrational mass-disorder model. 
We introduce the electronic width parameter $\mathcal{W}= \Delta m |\omega^2|$ for the width of the electronic disorder distribution. This allows us to estimate the vibrational mass disorder phase boundary from the electronic on-site disorder phase boundary \cite{BulSK87}.

In Fig.\ \ref{phase}(a) we show the estimated mobility edges for the case of vibrational mass disorder. The phase diagram is intriguing in many respects. We first note that the region for $\omega^2\geq 6$ corresponds to the $E\leq 0$ region in the Anderson model and similarly $\omega^2\leq 6$ is associated with $E\geq 0$. The much studied centre of the band at $E=0$ for the Anderson model becomes the much less distinct $\omega^2=6$ line in Fig.\ \ref{phase}(a). For $\omega^2\leq 6$, we see that much of the extended phase belongs to the region of possible negative masses with $\Delta m > 2$. Hence this region corresponds to a well studied counterpart in the Anderson case. Furthermore, the $E\geq 0$ region also extends into negative values of $\omega^2$ and gets transformed into a much reshaped form for $\omega^2<0$. The particular form of this puddle of extended states, towards the $\omega^2=0$ axis, is driven by the so-called re-entrant behaviour for the Anderson model \cite{BulSK87,GruS95}. Similarly, the re-entrant behaviour at $\omega^2>12$ can be traced to the corresponding re-entrant shape of the mobility edge at $E \lesssim -6$. As we will show, these extraordinary mobility edges and hence the phase diagram for the mass disorder case are indeed confirmed by our direct high-precision numerics.

For spring disorder, \eqref{eq-purespring} corresponds to a disorder distribution consisting of the sum of $6$ independently chosen random numbers. Even when each $k_{ij}$ is chosen according to the uniform distribution as above, the resulting distribution of $\epsilon_i$ has not previously been studied for the Anderson model --- although studies with pure hopping disorder exist \cite{SouE81} --- such that there are no phase diagrams to compare with. We have determined this phase diagram using our high-precision TMM.

In the TMM a quasi-onedimensional bar with fixed cross section $M\times M$ for lengths $L \gg M$ is considered. Equation (\ref{eq-dynmat}) is then rearranged into a form in which the amplitudes of vibration for a cross-sectional sheet can be calculated solely from those in the preceding cross-sectional sheet. If we denote the $i$-th $M\times M$ sheet of displacements by $U_i$, Eq.\ \eqref{eq-dynmat} can be expressed recursively as $(U_{i+1}, U_i)  = {\mathbb{T}_i} \cdot (U_i, U_{i-1})$ with suitably defined transfer matrix $\mathbb{T}_i$ \cite{PinSWR11a}. The standard TMM for the Anderson problem \cite{KraM93} is then used to calculate the Lyapunov exponent $\lambda_M$ of the mapping via $\mathbb{T}_i$'s and the reduced decay length $\Lambda_M=\lambda_M/M$.

We performed TMM calculations at various values of $\Delta m$, $\Delta k$ indicated in the phase diagrams in Fig.\ \ref{phase}. For every disorder value, we calculate $\Lambda_M$ for a range of frequencies and system widths $M= 6, 8, 10$ and $12$ to a relative error of $0.1\%$. The transition is initially estimated as the frequency where the $M=12$ and $M=10$ lines cross. The error shown is the difference with respect to the frequencies where the $M=12$ and the $M=6$ data cross. We see from Fig.\ \ref{phase}(a) that these rough estimates are in excellent agreement with the phase boundaries as established from the analogy with the Anderson transition. In Fig.\ \ref{phase}(b) the same method is used to obtain the phase diagram for the spring disorder. In addition, we performed studies at larger system sizes, up to $M=20$ with $0.1\%$ error, for the delocalization-localization transitions at three representative regions in the phase diagrams indicated by open diamonds in Fig.\ \ref{phase}. We found clear transitions from extended behaviour, with increasing $\Lambda_M$ values for increasing $M$, to localized behaviour, where $\Lambda_M$ decreases when $M$ increases. The transition frequencies obtained are in excellent agreement with the estimates by the method described above. 
%

For spring disorder we see that in the ``central region'' around $\omega^2=6$ states remain extended up to the largest considered spring disorder $\Delta k=10$. This is similar to the electronic case with pure hopping disorder \cite{SouE81} where even very strong hopping disorder does not lead to complete localization close to $E=0$. Whether the re-entrant behaviour of the phase border above the instability line for $\omega^2>8$ is genuine, remains to be determined by higher-precision calculation. We note that it coincides with the vanishing of the boson peak. Such a re-entrant regime is absent in the off-diagonal Anderson system. For the spring-disorder model it could signify a combined effect of localization and instability. For $\omega^2<0$ (and $\Delta k \gtrsim 4$), we observe an even larger area of extended states than for mass disorder. The localization-delocalization transition on the $\omega^2<0$ side is very similar to that observed in the instantaneous-normal mode spectra. The significance of the delocalized unstable modes to the energy landscape of a liquid remains to be discussed. 

We find that both for mass and spring disorder, the $\omega=0$ hydrodynamic mode remains extended regardless of the disorder strength. This is in agreement with previous studies in one- and two-dimensional systems \cite{LudSTE01}. What is also common to both mass and spring disorder is the observation of very strong shifts of the crossing points of $\Lambda_M$ when changing $M$. Such a behaviour is to be expected, however, since we are effectively dealing with transition regions in the vicinity of the band edges where density-of-states effects can dominate the scaling. This is again similar to the situation for the electronic case where the transition at the mobility edges for $E\neq 0$ is also more difficult to study \cite{MacK81,KraBMS90}.

We have also computed the density of states $g(\omega)$ of our models by directly diagonalizing the dynamical matrix. We have divided $g(\omega)$ by $\omega^2$ in order to detect maxima which correspond to ``boson peaks'', which have been discussed in the literature. In the stable regime below the boson maxima the vibrational excitations are essentially wave-like excitations.  For mass disorder we find such maxima within the whole phase diagram between $\omega=0$ and the mobility edge as shown in Fig.\ \ref{phase}(a). For spring disorder the boson peaks disappear slightly above the instability line $\Delta k=2$ (cp.\ Fig.\ \ref{phase}(b)). We have some evidence from our density-of-states calculations and from the form of the wave functions that indeed wave-like excitations persist in the instable region of the mass-disorder model, whereas there is no evidence for wave-like excitations in the spring disorder model for $\Delta k > 2.5$. This striking difference may be due to the fact that for mass disorder the disorder fluctuations are suppressed by a factor $\omega^2$, leaving a slightly disturbed spectrum of the simple cubic lattice in the small-$\omega$ regime. For spring disorder the band character is obviously destroyed already for small values of $\Delta k$.

\begin{table}[tb]
\begin{tabular}{cccccc}
\hline
\hline
$\Delta m$ & $M$ & $\omega^2$ & $\omega^2_c$ & $\nu$ & $\Gamma_q$\\
\hline
1.2 & 8 to 20 & $12.15$ to $13.1$ & $12.68 \pm 0.06$ & $1.57 \pm 0.14$ & 0.84\\
4 & 8 to 20 & $3.75$ to $4.25$  & $4.13 \pm 0.04$ & $1.57 \pm 0.08$ & 0.99\\
9 & 8 to 20 & $-1.65$ to $-1.5$ & $-1.62 \pm 0.04$ & $1.57 \pm 0.41$ & 0.87\\
\hline
\hline
$\Delta k$ & $M$ & $\omega^2$ & $\omega^2_c$ & $\nu$ & $\Gamma_q$\\
\hline
1 & 10 to 20 & $12.48$ to $12.6$ & $12.527 \pm 0.003$ & $1.58 \pm 0.05$ & 0.62\\
10 & 6 to 16 & $18.8$ to $20.3$  & $19.75 \pm 0.05$ & $1.51 \pm 0.08$ & 0.84\\
7 & 8 to 20 & $-3.5$ to $-2.75$  & $-3.33 \pm 0.11$ & $1.59 \pm 0.29$ & 0.51\\
\hline
\end{tabular}
\caption{Values of critical parameter $\omega_\text{c}^2$ and $\nu$ for spring and mass disorder computed from FSS performed in the given $M$ and $\omega^2$ ranges. The goodness-of-fit parameter $\Gamma_q$ is also shown for each fit. Errors denote the standard error-of-mean estimates.}
\label{tab-criticalparameter-w2}
\end{table}
We turn now to a discussion of our high-precision determination of the critical parameters. In order to ascertain the existence of a divergent correlation length $\xi(\omega^2) \propto |\omega^2 - \omega^2_c|^{-\nu}$ at $\omega^2_\text{c}$ with critical exponent $\nu$, we need to proceed, as usual in the electronic case, via a finite-size scaling (FSS) procedure \cite{SleO99a}. The FSS includes corrections to scaling which (i) account for the nonlinearities of the $\Delta m$, $\Delta k$ dependence of the scaling variables (relevant scaling) and (ii) for the shift of the point at which the $\Lambda_M(\omega^2)$ curves cross (irrelevant scaling). This analysis is by now standard and we refer to the literature for details of when fits are acceptable as stable and robust as well as for error estimates via Monte-Carlo approaches \cite{SleO99a,RodVSR10}. Details for the chosen expansions in the present case can be found in Ref.\ \cite{PinSWR11a}. In Tab.\ \ref{tab-criticalparameter-w2}, we show the results for the high-precision FSS analysis at the $6$ representative disorders. We find that in all cases, a consistent, robust and stable fit can be found with quality-of-fit parameter $\Gamma_q$ larger than $0.1$.
As our results show, we find that the critical exponents for both mass and spring disorder in the stable, unstable and $-\omega^2$ regions of the phase diagram agree with each other within the error estimates. In addition, they agree equally well with current estimates of the corresponding exponent for the Anderson model of localization \cite{MacK81,Mac94,SleO99a,RodVSR10}. Therefore we conclude that the scalar model of lattice vibrations studied here falls into the universality class of the Anderson transition \cite{AkiO98,MonG10}. 

\begin{figure}[tb]
\centering
 (a)\includegraphics[width=0.4\columnwidth,clip]{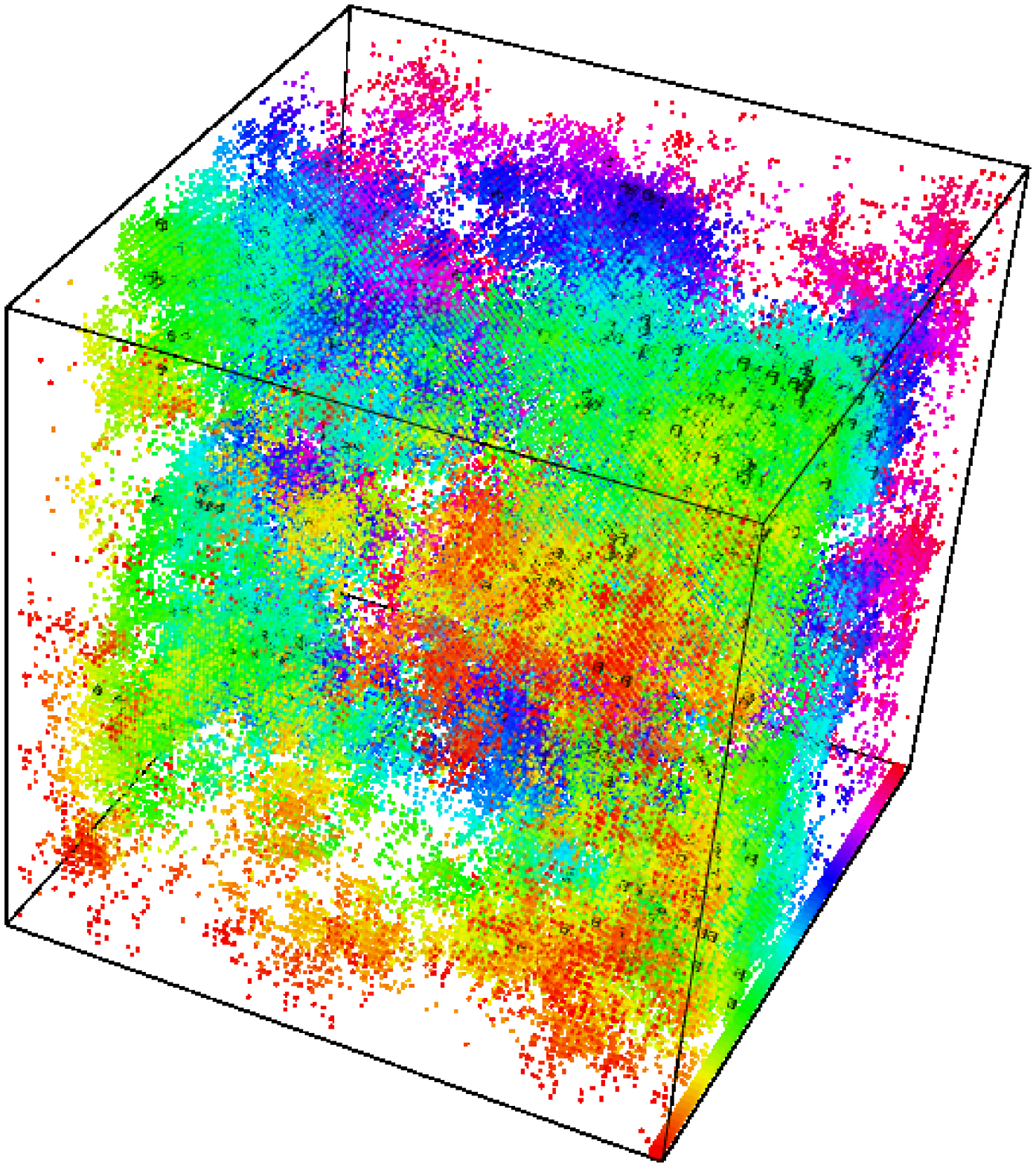} 
 (b)\includegraphics[width=0.4\columnwidth,clip]{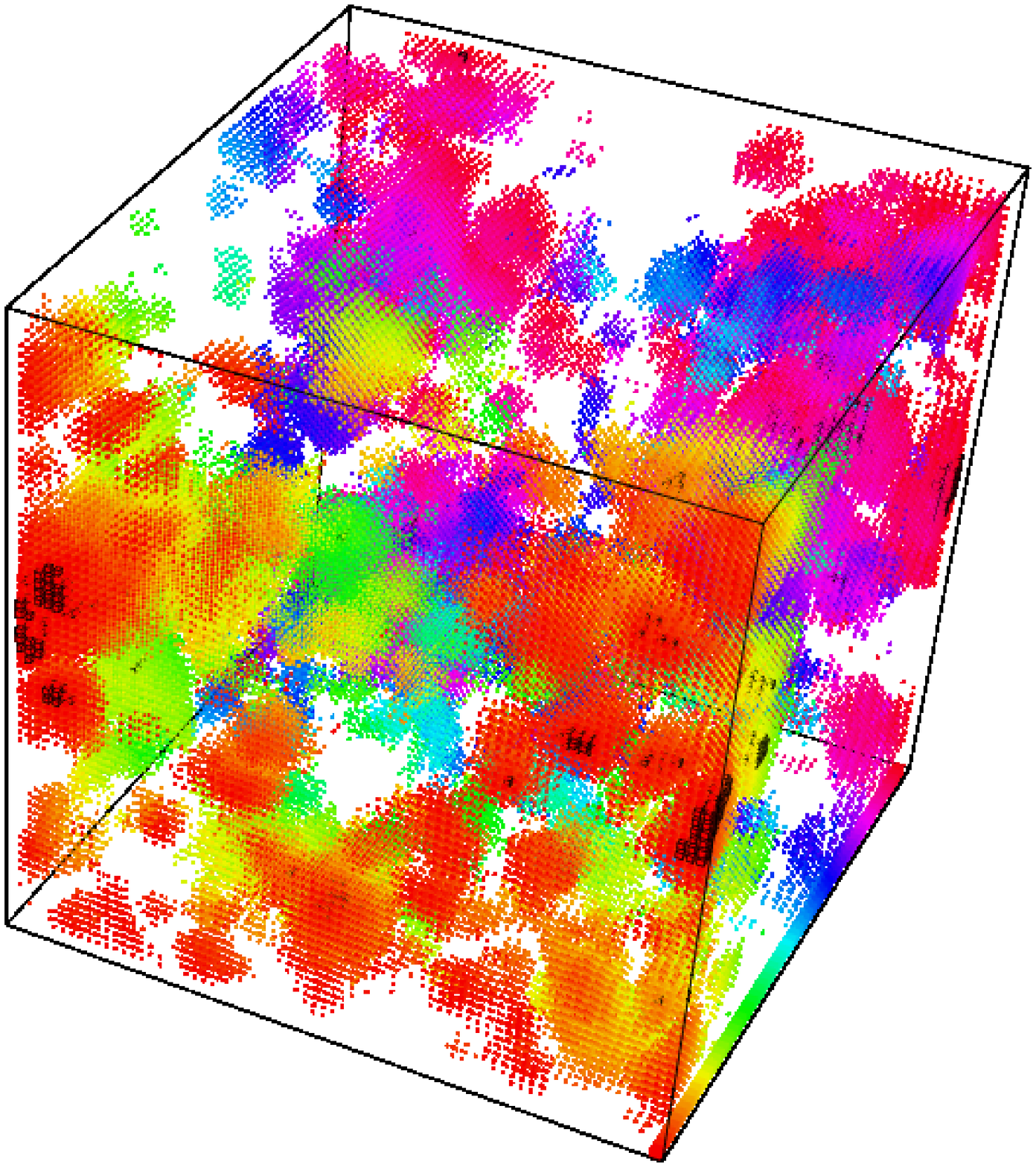}
\caption{(Color online) Schematic representation of critical amplitude distributions obtained from exact diagonalization for (a) $\Delta m=4$ at $\omega^2 = 4.134$ and (b) $\Delta k=1$ at $\omega^2 = 12.526$. All sites with $\frac{u_{x,y,z} L^{-3}}{\sum_{x,y,z} u_{x,y,z}} >1$ are shown as small cubes and those with black edges have $\frac{u_{x,y,z} L^{-3}}{\sum_{x,y,z} u_{x,y,z}} > \sqrt{1000}$. The color scale distinguishes between different slices of the system along the axis into the page.\label{fig-states}}
\end{figure}

In Fig.\ \ref{fig-states} we have represented critical amplitude distributions of the states marked with black circles in Fig.\ \ref{phase}. We see that the structure for mass disorder is quite different from that of the spring disorder. An evaluation of the multifractal properties of these states might probably exhibit different spectra for the two models.

In conclusion, our results show that a disordered scalar phonon model exhibits all the rich features of the Anderson localization-delocalization transition. While the critical exponents are universal and of Anderson type the mass-disorder and spring-disorder models exhibit completely different localization phase diagrams. The re-entrant behaviour of the mass-disorder system --- inherited from the Anderson model with on-site disorder --- is present both on the stable and unstable side of the phase diagram. The spring-disorder phase diagram is dominated by delocalized states. Localized states exist on both sides near the band edges. In the stable regime $\Delta m<2, \Delta k<2$ localized states exist only at the upper band edge in agreement with earlier investigations \cite{JSS83,SchDG98}.

\acknowledgments
SDP and RAR are grateful for discussions with E Parker, A Rodriguez-Gonzalez and T Whall. We thankfully acknowledge EPSRC (EP-F040784-1) and the EC ``Nanofunction'' Network of Excellence for financial support.


%
%

\end{document}